# Heterogeneous networks for phase-sensitive engineering of optical disordered materials

Seungmok Youn[1], Kunwoo Park[1], Ikbeom Lee[1], Gitae Lee[1], Namkyoo Park[2†], and Sunkyu Yu[1*]

[1]Intelligent Wave Systems Laboratory, Department of Electrical and Computer Engineering, Seoul National University, Seoul 08826, Korea

[2]Photonic Systems Laboratory, Department of Electrical and Computer Engineering, Seoul National University, Seoul 08826, Korea

E-mail address for correspondence: [†]nkpark@snu.ac.kr, [*]sunkyu.yu@snu.ac.kr

**Abstract**

Heterogeneous networks provide a universal framework for extracting subsystem-level features of a complex system, which are critical in graph colouring, pattern classification, and motif identification. When abstracting physical systems into networks, distinct groups of nodes and links in heterogeneous networks can be decomposed into different modes of multipartite networks, allowing for a deeper understanding of both intra- and inter-group relationships. Here, we develop heterogeneous network modelling of wave scattering to engineer multiphase random heterogeneous materials. We devise multipartite network decomposition determined by material phases, which is examined using uni- and bi-partite network examples for two-phase multiparticle systems. We show that the directionality of the bipartite network governs the phase-sensitive alteration of microstructures. The proposed modelling enables a network-based design to achieve phase-sensitive microstructural features, while almost preserving the overall scattering response. With examples of designing quasi-isoscattering stealthy hyperuniform materials, our results provide a general recipe for engineering multiphase materials for wave functionalities.

# Introduction

Random heterogeneous materials[1]—materials composed of multiple phases with statistically characterized microstructural distributions—have provided extended design freedom in manipulating optical[2-5], electrical[3,6], and mechanical[7-9] properties. Although a traditional viewpoint had separated the unique characteristics of such random media from those of ordered ones—for example, wave localization[10] and gap annihilation[11] in disordered materials, as opposed to extended Bloch modes and bandgaps in crystalline structures[12]—engineering disorder has recently attracted attention for devising advanced functionalities unattainable by traditional classifications. Notable examples include crystal-like gap dynamics achieved with disordered stealthy hyperuniformity (SHU)[13-17], topological band phenomena in disordered materials[18,19], and dynamical tunnelling in deformed resonators[20,21].

When implementing functional materials, the microstructural characteristics of each material phase are critical because the target function often originates from a subset of the phases. For example, when devising nonlinear devices, nonlinear materials are usually embedded within linear materials with different material parameters[22]. In exploring non-Hermitian physics using optical testbeds[23], the unique arrangement of materials possessing distinct amplification or dissipation enables unconventional wave phenomena. Therefore, material-phase-selective engineering of disorder is critical for fully exploiting the extended degrees of freedom in random heterogeneous materials to achieve advanced functionalities.

Here, we propose the concept of heterogeneous networks to model and design optical scattering from multiphase random heterogeneous materials. Under the Born approximation, we define the weight of edges and node degrees, which reflect the contribution of each wave interference to the overall scattering. Using this network modelling, we decompose the network to multipartite subnetworks, which represent the wave-based connectivity in and between material subphases. By developing evolutionary material design using network parameters, we examine two-phase materials, revealing the impact of the directionality of the bipartite network on phase-sensitive engineering of material microstructures. The proposed method enables the inverse design of materials yielding almost identical scattering responses in the reciprocal space of interest, while differentiating those in the complementary space.

## Results

**Heterogeneous networks of scattering**

We investigate the heterogeneous network modelling of scattering from an inhomogeneous $N$-particle material (Fig. 1a). When the optical potential perturbation induced by the $j$th particle is $V_j(\mathbf{r})$ bounded around the particle, the material is characterized by the potential $V(\mathbf{r}) = \sum_j V_j(\mathbf{r})$. Under the weak scattering assumption and far-field observation, the scattering intensity from the incident wavevector $\mathbf{k}_\mathrm{I}$ becomes $I_N(\mathbf{k}) = |\int V(\mathbf{r})e^{-i\mathbf{k}\cdot\mathbf{r}}d\mathbf{r}|^2$, where $\mathbf{k} = \mathbf{k}_\mathrm{S} - \mathbf{k}_\mathrm{I}$ is the momentum shift for the scattering wavevector $\mathbf{k}_\mathrm{S}$. The structure factor of the material is defined as $S(\mathbf{k}) = I_N(\mathbf{k})/[NI_\mathrm{avg}(\mathbf{k})]$, where $I_\mathrm{avg}(\mathbf{k}) = (1/N)\sum_j |\int V_j(\mathbf{r})e^{-i\mathbf{k}\cdot\mathbf{r}}d\mathbf{r}|^2$ is the average of scattering intensity from a single particle[24].

By generalizing the scattering network modelling for a single-phase, point-particle material[25], the edge weight between the $p$th and $q$th particles for a heterogeneous scattering network is defined as (Supplementary Note S1):

$$w_{p,q}^\mathrm{K} = \frac{1}{\Omega_\mathrm{K}} \int_\mathrm{K} \frac{\mathrm{Re}(v_p(\mathbf{k})v_q^*(\mathbf{k}))}{I_\mathrm{avg}(\mathbf{k})} d\mathbf{k} \qquad (1)$$

where $v_p(\mathbf{k})$ and $v_q(\mathbf{k})$ are the Fourier transforms of $V_p(\mathbf{r})$ and $V_q(\mathbf{r})$, respectively, $\mathbf{K}$ is the region of interest in the reciprocal space, and $\Omega_\mathbf{K}$ is the volume of $\mathbf{K}$. The edge weight characterizes the contribution of the $(p,q)$th particle pair to the entire scattering over $\mathbf{K}$. The material can be interpreted as a fully-connected weighted network with real-valued edge weights $w_{p,q}^\mathbf{K}$. We note that the node degree of a particle is defined as $w_p^\mathbf{K} = \sum_{q \neq p} w_{p,q}^\mathbf{K}$, which bridges our network model to scattering, as $\langle S(\mathbf{k}) \rangle_\mathbf{K} = 1 + (1/N)\sum_p w_p^\mathbf{K}$, where $\langle ... \rangle_\mathbf{K}$ represents the average over $\mathbf{K}$.

In contrast to the scattering network model for a single-phase material[25], the generalized model can be decomposed into a set of multipartite networks to characterize a multi-phase material. The multipartite networks of the network are distinguished from one another by the material-phase-dependent participating nodes and edges. For example, without any loss of generality, consider a material composed of two-phase particles (Fig. 1b), where the phases 'A' and 'B' denote different material or structural features. The corresponding heterogeneous scattering network includes the edges connecting the A-A, B-B, and A-B phases, which lead to three multipartite networks (Fig. 1c-e): two unipartite networks ('A net' in Fig. 1c and 'B net' in Fig. 1e), and a bipartite network ('AB net' in Fig. 1d). According to this network decomposition, the edge weights $w_{p,q}^\mathbf{K}$ are classified into each multipartite network, as ($p \in A$, $q \in A$) and ($p \in B$, $q \in B$) for the unipartite networks, and ($p \in A$, $q \in B$) for the bipartite network.

In the proposed model, the structure factor, and the corresponding scattering component, are decomposed into that of each multipartite network: $\langle S(\mathbf{k}) \rangle_\mathbf{K} = 1 + \langle S_{AA}(\mathbf{k}) \rangle_\mathbf{K} + \langle S_{BB}(\mathbf{k}) \rangle_\mathbf{K} + \langle S_{AB}(\mathbf{k}) \rangle_\mathbf{K}$, where $S_{XY}(\mathbf{k})$ denotes the contribution from the XY phase (X, Y $\in$ {A, B}. Supplementary Note S2). Therefore, the network decomposition allows for characterizing the impact of each material phase and their interactions on the entire scattering event.

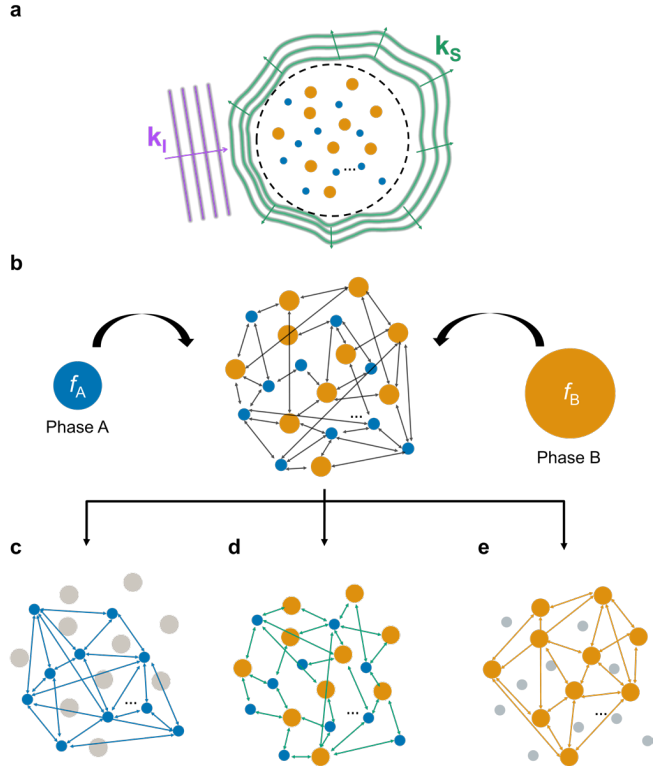

**Figure 1. Heterogeneous scattering networks**. **a**, Wave scattering in a two-phase *N*-particle material. The colour and radius of each particle denote its material phase and distinct geometry, respectively. **b**, Heterogeneous scattering network, modelling particles as nodes and wave interferences from pairs of particles as edges. **c-e**, Multipartite networks characterized by material phases: unipartite A net (**c**), bipartite AB net (**d**), and unipartite B net (**e**). All edges possess different weight values according to Eq. (1). Although the heterogeneous scattering network and its decomposed multipartite networks are originally fully connected, only a portion of the entire edges are illustrated for visibility.

**Designing quasi-isoscattering**

In network science, exploring alternative topologies that exhibit similar properties is a key research theme across multiple subfields, as demonstrated by the development of isospectral[26] and equienergetic[27] graphs and by the design of graph kernels[28]. In a similar context, identifying distinct scattering networks that exhibit nearly identical scattering responses enables an inverse design strategy within the framework of engineering disorder[29] in photonics—designing materials with different microstructural statistics that permit selective control over specific optical responses. In our example, we focus on achieving a specified scattering response of two-phase materials. The response is characterized by a suppressed structure factor within the reciprocal-space range $\mathbf{K} = \{\mathbf{k} \mid |\mathbf{k}| < k_{th}\}$, which corresponds to the finite-sample implementation of SHU[17,29,30]. The target two-phase material is composed of the phases 'A' and 'B', where the two phases are distinguished by their strength of potential perturbation. To ensure a fair comparison, we preserve the total amount of potential perturbations in the resulting materials.

We devise the evolutionary design to update the two-phase heterogeneous network (Fig. 2a), which includes two processes: first, the Bernoulli-random selection of the material phase of a new particle, and second, the estimation of its position using the cost function. To define the cost function, we utilize $\langle S(\mathbf{k}) \rangle_\mathbf{K} = 1 + \langle S_{AA}(\mathbf{k}) \rangle_\mathbf{K} + \langle S_{BB}(\mathbf{k}) \rangle_\mathbf{K} + \langle S_{AB}(\mathbf{k}) \rangle_\mathbf{K}$, allowing for decomposing node degrees into multipartite network parameters:

$$w_{p,u}^\mathbf{K} \equiv \sum_{\substack{q \neq p, \\ p,q \in A \text{ or } p,q \in B}} w_{p,q}^\mathbf{K}, \quad w_{p,b}^\mathbf{K} \equiv \sum_{\substack{(p \in A, q \in B) \text{ or} \\ (p \in B, q \in A)}} w_{p,q}^\mathbf{K}, \qquad (2)$$

where the subscripts 'u' and 'b' correspond to unipartite and bipartite networks. Using Eq. (2), we obtain $\langle S_{AA}(\mathbf{k}) \rangle_\mathbf{K} + \langle S_{BB}(\mathbf{k}) \rangle_\mathbf{K} = (1/N)\sum_p w_{p,u}^\mathbf{K}$ and $\langle S_{AB}(\mathbf{k}) \rangle_\mathbf{K} = (1/N)(\sum_p w_{p,b}^\mathbf{K})$. The cost function $\rho_n^\mathbf{K}(\mathbf{r})$ at the evolution step $n$ is then defined as a weighted sum of the newly generated node degrees, where $\mathbf{r}$ denotes the candidate position, as follows:

$$\rho_n^\mathbf{K}(\mathbf{r}) = \lambda_u w_{n,u}^\mathbf{K}(\mathbf{r}) + \lambda_b w_{n,b}^\mathbf{K}(\mathbf{r}), \qquad (3)$$

where the coefficients $\lambda_u$ and $\lambda_b$ manipulate the relative impact of the intra- and inter-edges over material phases during the design process, respectively.

Based on the proposed design strategy, we design a family of quasi-isoscattering two-phase materials—that is, materials that induce nearly identical scattering responses[31] over **K**—in a two-dimensional (2D) plane. To examine the effect of inhomogeneity, we compare the designs with varying material-phase form factor ratio $f_B/f_A$, where $f_A$ and $f_B$ denote a scattering form factor of a phase 'A' and phase 'B' particle, respectively (Supplementary Note S3). Because $f_A$ and $f_B$ represent the node characteristics in scattering networks, our problem corresponds to exploring quasi-isoscattering networks under the variation of nodes. Figure 2b shows that the suppressed scattering over **K** is well preserved despite varying inhomogeneity, yielding a family of finite SHU materials of which the level of scattering suppression is nearly identical to that of a finite square lattice (dashed line). This SHU condition is also evident in the reciprocal-space plots shown in the insets of Fig. 2b, which simultaneously reveal distinct short-wavelength (that is, large |**k**|) responses for each inhomogeneity. We note that the short-wavelength discrepancy originates from different microstructures of two-phase materials, as shown in Fig. 2c, exhibiting the impacts of the multipartite network edges. In extremely inhomogeneous conditions ($f_B/f_A \approx 0$), the unipartite network corresponding to the phase with the larger form factor ('A net') dominates the overall effect, while the unipartite network of the other phase ('B net') and the bipartite network ('AB net') merely contribute. As the materials are homogenized ($f_B/f_A \to 1$), the contributions of the unipartite nets converge, and the effect of the bipartite net increases.

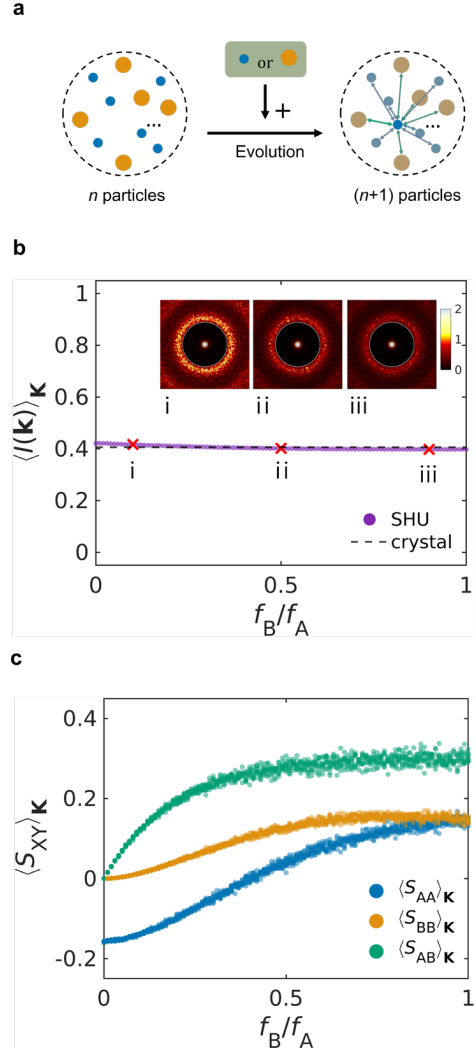

**Figure 2. Evolutionary design. a**, Evolution of a two-phase *n*-particle material at the $(n+1)^{th}$ evolution step. Binomial phase selection is followed by optimal position estimation. **b**, The scattering $\langle I(\mathbf{k})\rangle_\mathbf{K}$ of designed materials, where $\mathbf{K} = \{\mathbf{k} \mid |\mathbf{k}| < 0.5 k_c\}$ and $I(\mathbf{k}) = I_N(\mathbf{k})/N$. $k_c$ is the wavenumber defined by the characteristic length, where $k_c = \pi N^{1/2}/R$, where $R$ is the radius of the multiparticle system in real space. The dashed line indicates $\langle I(\mathbf{k})\rangle_\mathbf{K}$ of a square crystal consisting of 3,001 identical particles, where each particle is designed to maintain the overall potential perturbation. Red markers denote the phase inhomogeneity conditions $f_B/f_A = 0.1$ (i), $f_B/f_A = 0.5$ (ii), and $f_B/f_A = 0.9$ (iii), respectively, and the insets denote the corresponding reciprocal-space profiles of ensemble-averaged $I(\mathbf{k})$. **c,** The scattering contributions of edges in different phases. $S_{XY}(\mathbf{k})$ denotes the contribution from the edges in the XY phase (X, Y ∈ {A, B}). Weight

coefficients are set to be $\lambda_u = \lambda_b = 1$. In b and c, each material consists of 3,000 particles, and 10 realizations are generated for each phase inhomogeneity condition.

**Network connectivity.** Although the results in Fig. 2 unveil the possibility of tailoring material microstructures while preserving scattering, the underlying mechanism—from a network perspective—may originate from node manipulation or the resulting changes in network connectivity. To isolate the impact of connectivity, which solely governs particle distributions, we examine the evolution process while maintaining the inhomogeneity constant as $f_B/f_A = 0.5$. In particular, we alter the ratio of weighting coefficients, $\lambda_b/\lambda_u$, to explore quasi-isoscattering networks composed of differentiated multipartite networks (Fig. 3a). The resulting network connectivity is investigated through the node degree distributions of subnetworks, which have been widely employed in characterizing network topology[32]. In examining the node degree distribution of the bipartite net, we decompose it into two distinct distributions according to the material phase of a node (Fig. 3b), which imposes directivity on the edges of the bipartite network in analysing its network connectivity. We emphasize that such directional bipartite networks operate as material microstructure identifiers, for example, describing the microstructure of phase A observed from phase B using $w_{p \in B,b}^K$ (see Supplementary Note S4 for details).

Figures 3c-f show the node degree distributions of unipartite networks (Fig. 3c,d) and directional bipartite networks (Fig. 3e,f) for different $\lambda_b/\lambda_u$. By controlling $\lambda_b/\lambda_u$, we can manipulate the impact of each multipartite network in achieving the target scattering response—the significant role of the unipartite network with $\lambda_b = 0$ (Fig. 3c,e) and of the bipartite network with $\lambda_b > \lambda_u$ (Fig. 3d,f)—which mainly possesses negative node degrees. Especially, when we neglect the change of bipartite networks during the design ($\lambda_b/\lambda_u = 0$), we achieve strong agreement of directional bipartite networks (Fig. 3e), implying the same connectivity between different phases. In contrast,

the bipartite network becomes apparently directional by increasing $\lambda_b/\lambda_u$ (Fig. 3f), indicating that the microstructures of phases A and B are significantly differentiated while maintaining the overall scattering response.

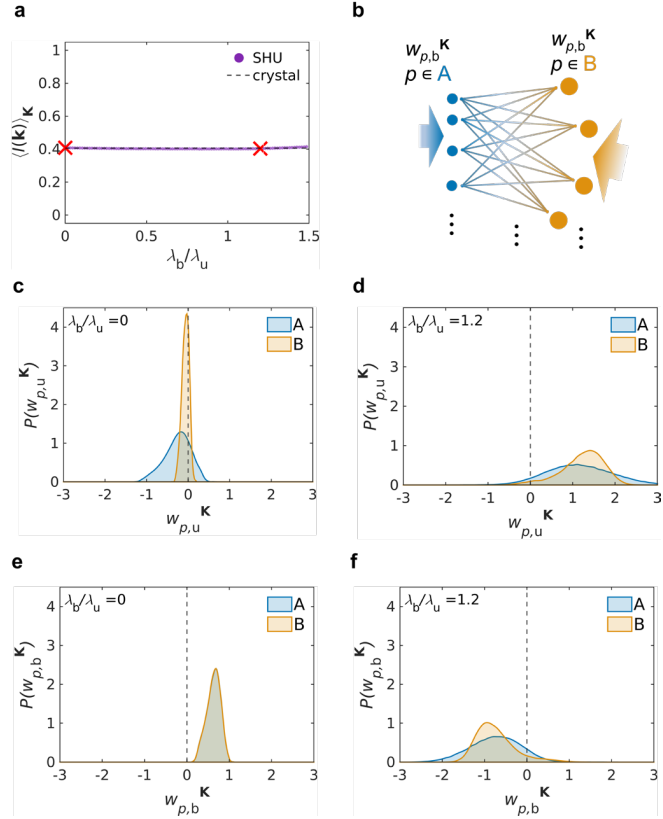

**Figure 3. Degree distributions of multipartite networks. a,** Variation of $\langle I(\mathbf{k})\rangle_\mathbf{K}$, where $\mathbf{K} = \{\mathbf{k} \mid |\mathbf{k}| < 0.5k_c\}$, by the ratio of the weighting coefficients $\lambda_b/\lambda_u$. Red crosses denote $\lambda_b/\lambda_u = 0$ and 1.2. The dashed line shows $\langle I(\mathbf{k})\rangle_\mathbf{K}$ of a square crystal consisting of 3,001 identical particles. **b,** Directional graph topology of a bipartite net. Node degree distributions are obtained for the node phases 'A' and 'B' separately. **c-f,** Node degree distributions of unipartite (**c,d**) and bipartite (**e,f**) networks for the configurations of $\lambda_b/\lambda_u = 0$ (**c,e**) and 1.2 (**d,f**). All distributions are generated from 100 realizations of materials, each consisting of 3,000 particles.

**Phase-sensitive microstructures.** Given the relationship between scattering network weights and structure factors, the observed directionality in the bipartite scattering network suggests that material phases A and B possess distinct microstructures despite the overall quasi-isoscattering across $\mathbf{K}$ as shown in Fig. 3f. To rigorously validate this prediction, we analyse the microstructural statistics of each material phase for different configurations of $\lambda_b/\lambda_u = 0$ (Fig. 4a-c), $\lambda_b/\lambda_u = 1.0$ (Fig. 4d-f), and $\lambda_b/\lambda_u = 1.2$ (Fig. 4g-i), where higher $\lambda_b/\lambda_u$ values correspond to more directional bipartite networks. We examine the microstructural statistics using the scaled number variance[33,34] $\sigma^2(R)/\mu(R)$, where $\sigma^2(R)$ and $\mu(R)$ denote the variance and mean, respectively, of the number of particles within a circular window of radius $R$. It is well known that while the Poisson uncorrelated disorder yields a constant $\sigma^2(R)/\mu(R)$, generalized ordered material phases, including crystals and SHU, exhibit a $1/R$ decay in the thermodynamic limit. For comparison with our design, we include $\sigma^2(R)/\mu(R)$ for crystal, SHU, and Poisson materials in finite real space in Supplementary Note S5.

As expected from the undirected bipartite network in Fig. 3e, $\lambda_b/\lambda_u = 0$ leads to the identical microstructural statistics across material phases; when compared with the statistics of the overall heterogeneous material of A∪B—which shows the SHU-like suppressed variance—both phases A and B show almost identical variances as shown in Fig. 4b. Consequently, the heterogeneous material obtained with $\lambda_b/\lambda_u = 0$ can be regarded as a simple aggregation of SHU materials in different phases. In contrast, the design with larger $\lambda_b/\lambda_u$ leads to highly phase-sensitive alterations in microstructures. Specifically, despite the quasi-isoscattering over $\mathbf{K}$ (Figs. 4c,f,i), and also the similar variance suppression (black lines in Figs. 4b,e,h) of the entire material, Figs. 4e and 4h reveal markedly distinct microstructural statistics for phases A and B, showing that scaled number variances of phase B significantly deviate from the SHU-like variances of phase A. At $\lambda_b/\lambda_u = 1.0$, phase B exhibits the Poisson-like variance (Fig. 4e) when compared with Supplementary Fig. S2. Interestingly, the design with highly directional bipartite network using $\lambda_b/\lambda_u = 1.2$ leads to greatly enhanced variance at a specific $R$ (Fig. 4h), which represents the clustering of particles described in Fig. 4g. The observed microstructural statistics yields the alteration of short-range density fluctuations (Fig. 4c,f,i), allowing for the length-scale-dependent engineering of light scattering through the phase-selective engineering of random heterogeneous materials.

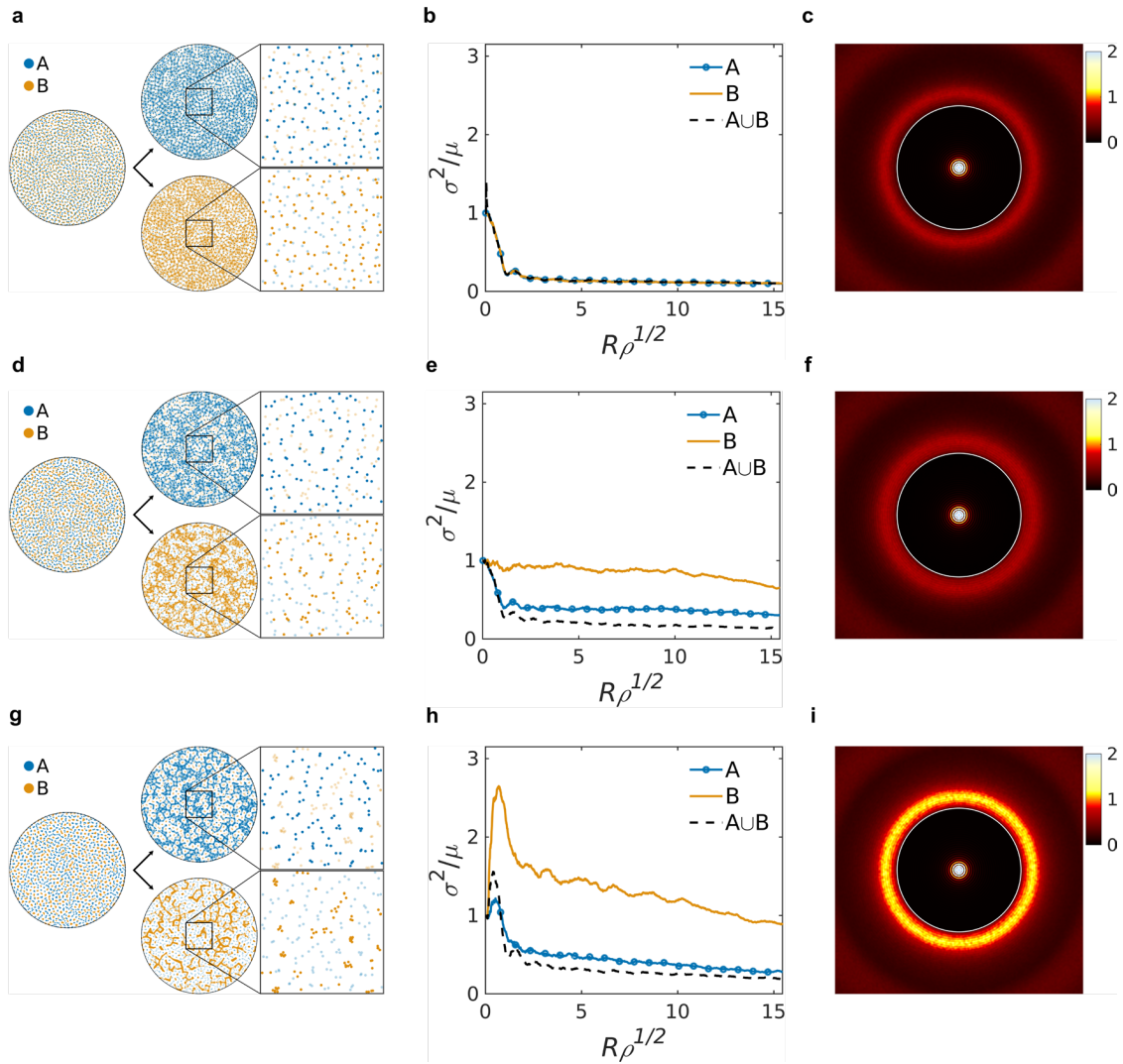

**Figure 4. Scaled variances.** Designs with $\lambda_b/\lambda_u = 0$ (**a-c**), 1.0 (**d-f**), and 1.2 (**g-i**). **a,d,g,** Visualized unipartite nets. Left panels demonstrate the sample materials and subgraphs of unipartite networks. 2,000 edges of each A net and B net, with the least edge weights, are illustrated, while thickness and transparency reflect their weights. Right panels show magnified particle arrangements sampled with squares of area $200/\rho$, emphasizing phase A particles (upper box) and phase B particles (lower box), respectively, where $\rho$ denotes the total particle number density. Each material consists of 3,000 particles. **b,e,h,** Scaled number variances of particles belong to each phase as a function of scaled window radius $R\rho^{1/2}$. The statistics are obtained from 1,000 realizations each. **c,f,i,** Reciprocal-space profiles of ensemble-averaged scattering intensity $I(\mathbf{k})$.

## Discussion

We have proposed the heterogeneous scattering networks to model multi-phase optical disordered materials. By examining multipartite scattering networks for two-phase multiparticle materials, we devised the evolutionary design method for achieving an isoscattering family of materials. By manipulating the connectivity of subnetworks, we observed the designed alteration of network parameters and the consequent change of material microstructures. The resulting materials include both multi-hyperuniform[35,36] and effectively hyperuniform[37,38] realizations, especially displaying the deterministic engineering of short-range density fluctuations using the directionality of bipartite networks.

Reviewing numerous achievements with multipartite networks, including the characterization of network coefficients[39], community detection techniques[40], and the classification of entangled quantum states[41,42], we envisage that the presence of multiple subnetworks provides additional degrees of freedom for engineering light-matter interactions—not only by utilizing weighting coefficients $\lambda_b$ and $\lambda_u$, , but also defining cost functions that incorporate various preferential attachment schemes, which are key principles of evolving networks[43].

The concept of a heterogeneous scattering network can be extended to broader types of potential perturbations. While the current proposal can be applied beyond point-particle approximation under the Born approximation, the generalization to the multiple scattering regime is desired to cover strong perturbation regimes. The utilization of phase-selective engineering to active optical functionalities will also be a future research topic.

**Data availability**

The data that support the plots and other findings of this study are available from the corresponding author upon request.

## Code availability

All code developed in this work will be made available from the corresponding author upon request.

## Acknowledgements

We acknowledge financial support from the National Research Foundation of Korea (NRF) through the Basic Research Laboratory (No. RS-2024-00397664), Innovation Research Center (No. RS-2024-00413957), Young Researcher Programs (No. 2021R1C1C1005031), and Midcareer Researcher Program (No. RS-2023-00274348), all funded by the Korean government. This work was supported by Creative-Pioneering Researchers Program and the BK21 FOUR program of the Education and Research Program for Future ICT Pioneers in 2024, through Seoul National University. We also acknowledge administrative support from SOFT foundry institute.

## Author Contributions

S.Youn, N.P., and S.Yu conceived the idea. S.Youn developed the theoretical tool and performed the numerical analysis. K.P., I.L., and G.L. examined the theoretical and numerical analysis. S.Y. and N.P. supervised the findings of this work. All authors discussed the results and wrote the final manuscript.

## Competing Interests

The authors have no conflicts of interest to declare.

## Additional information

**Correspondence and requests for materials** should be addressed to N.P. or S.Y.

**Figure Legends**

**Figure 1. Heterogeneous scattering networks**. **a**, Wave scattering in a two-phase *N*-particle material. The colour and radius of each particle denote its material phase and distinct geometry, respectively. **b**, Heterogeneous scattering network, modelling particles as nodes and wave interferences from pairs of particles as edges. **c-e**, Multipartite networks characterized by material phases: unipartite A net (**c**), bipartite AB net (**d**), and unipartite B net (**e**). All edges possess different weight values according to Eq. (1). Although the heterogeneous scattering network and its decomposed multipartite networks are originally fully connected, only a portion of the entire edges are illustrated for visibility.

**Figure 2. Evolutionary design. a**, Evolution of a two-phase *n*-particle material at the $(n+1)^{th}$ evolution step. Binomial phase selection is followed by optimal position estimation. **b**, The scattering $\langle I(\mathbf{k})\rangle_K$ of designed materials, where $K = \{\mathbf{k} \mid |\mathbf{k}| < 0.5k_c\}$ and $I(\mathbf{k}) = I_N(\mathbf{k})/N$. $k_c$ is the wavenumber defined by the characteristic length, where $k_c = \pi N^{1/2}/R$, where $R$ is the radius of the multiparticle system in real space. The dashed line indicates $\langle I(\mathbf{k})\rangle_K$ of a square crystal consisting of 3,001 identical particles, where each particle is designed to maintain the overall potential perturbation. Red markers denote the phase inhomogeneity conditions $f_B/f_A = 0.1$ (i), $f_B/f_A = 0.5$ (ii), and $f_B/f_A = 0.9$ (iii), respectively, and the insets denote the corresponding reciprocal-space profiles of ensemble-averaged $I(\mathbf{k})$. **c,** The scattering contributions of edges in different phases. $S_{XY}(\mathbf{k})$ denotes the contribution from the edges in the XY phase (X, Y ∈ {A, B}). Weight coefficients are set to be $\lambda_u = \lambda_b = 1$. In b and c, each material consists of 3,000 particles, and 10 realizations are generated for each phase inhomogeneity condition.

**Figure 3. Degree distributions of multipartite networks. a,** Variation of $\langle I(\mathbf{k})\rangle_K$, where $K = \{\mathbf{k} \mid |\mathbf{k}| < 0.5k_c\}$, by the ratio of the weighting coefficients $\lambda_b/\lambda_u$. Red crosses denote $\lambda_b/\lambda_u = 0$ and 1.2. The dashed line shows $\langle I(\mathbf{k})\rangle_K$ of a square crystal consisting of 3,001 identical particles. **b,** Directional graph topology of a bipartite net. Node degree distributions are obtained for the node phases 'A' and 'B' separately. **c-f,** Node degree distributions of unipartite (**c,d**) and bipartite (**e,f**) networks for the configurations of $\lambda_b/\lambda_u = 0$ (**c,e**) and 1.2 (**d,f**). All distributions are generated from 100 realizations of materials, each consisting of 3,000 particles.

**Figure 4. Scaled variances.** Designs with $\lambda_b/\lambda_u = 0$ (**a-c**), 1.0 (**d-f**), and 1.2 (**g-i**). **a,d,g,** Visualized unipartite nets. Left panels demonstrate the sample materials and subgraphs of unipartite networks. 2,000 edges of each A net and B net, with the least edge weights, are illustrated, while thickness and transparency reflect their weights. Right panels show magnified particle arrangements sampled with squares of area $200/\rho$, emphasizing phase A particles (upper box) and phase B particles (lower box), respectively, where $\rho$ denotes the total particle number density. Each material consists of 3,000 particles. **b,e,h,** Scaled number variances of particles belong to each phase as a function of scaled window radius $R\rho^{1/2}$. The statistics are obtained from 1,000 realizations each. **c,f,i,** Reciprocal-space profiles of ensemble-averaged scattering intensity $I(\mathbf{k})$.

**Supplementary Information for "Heterogeneous networks for phase-sensitive engineering of optical disordered materials"**

Seungmok Youn[1], Kunwoo Park[1], Ikbeom Lee[1], Gitae Lee[1], Namkyoo Park[2†], and Sunkyu Yu[1*]

[1]Intelligent Wave Systems Laboratory, Department of Electrical and Computer Engineering, Seoul National University, Seoul 08826, Korea

[2]Photonic Systems Laboratory, Department of Electrical and Computer Engineering, Seoul National University, Seoul 08826, Korea

E-mail address for correspondence: [†]nkpark@snu.ac.kr, [*]sunkyu.yu@snu.ac.kr

**Note S1. Edge weights**

**Note S2. Contributions of multipartite networks to scattering**

**Note S3. Form factors of particles**

**Note S4. Bipartite networks as material microstructure identifiers**

**Note S5. Microstructural statistics of finite-size materials**

**Note S1. Edge weights**

When applying the potential perturbation $V(\mathbf{r}) = \sum_p V_p(\mathbf{r})$ to the structure factor, we obtain

$$S(\mathbf{k}) = \frac{\left|\int V(\mathbf{r})e^{-i\mathbf{k}\cdot\mathbf{r}}d\mathbf{r}\right|^2}{NI_{avg}(\mathbf{k})} = \frac{1}{NI_{avg}(\mathbf{k})}\left|\sum_p v_p(\mathbf{k})\right|^2, \tag{S1}$$

where $v_p(\mathbf{k})$ is the Fourier transform of $V_p(\mathbf{r})$, defined as $v_p(\mathbf{k}) = \int V_p(\mathbf{r})\exp(-i\mathbf{k}\cdot\mathbf{r})d\mathbf{r}$. To derive the edge weights for a heterogeneous scattering network, which are directly connected to wave scattering, we expand Eq. (S1) as

$$S(\mathbf{k}) = \frac{1}{NI_{avg}(\mathbf{k})}\sum_{p,q} v_p(\mathbf{k})v_q^*(\mathbf{k}) = 1 + \frac{2}{NI_{avg}(\mathbf{k})}\sum_{p\neq q,\{p,q\}} \text{Re}\{v_p(\mathbf{k})v_q^*(\mathbf{k})\}, \tag{S2}$$

where $\{p,q\}$ denotes the unordered pair except for $p = q$. The reciprocal-space average of $S(\mathbf{k})$ over $\mathbf{K}$ becomes

$$\langle S(\mathbf{k})\rangle_{\mathbf{K}} = 1 + \frac{2}{N}\sum_{p\neq q,\{p,q\}} \frac{1}{\Omega_{\mathbf{K}}}\int_{\mathbf{K}} \frac{\text{Re}\{v_p(\mathbf{k})v_q^*(\mathbf{k})\}}{I_{avg}(\mathbf{k})}d\mathbf{k}. \tag{S3}$$

Following the relationship between scattering and network parameters[1], the edge weights for our heterogeneous scattering network can be defined as Eq. (1) in the main text, which leads to $\langle S(\mathbf{k})\rangle_{\mathbf{K}} = 1 + (1/N)\sum_p w_p^{\mathbf{K}}$ where $w_p^{\mathbf{K}}$ denotes the node degree of the $p$th particle.

**Note S2. Contributions of multipartite networks to scattering**

The material phases of participating nodes differentiate the type of edges. Separately summing the weights of edges connecting A-A, B-B, and A-B material phases, the structure-factor contribution $S_{XY}(\mathbf{k})$ to $S(\mathbf{k})$ from the XY phase (X, Y $\in$ {A, B}) is defined as:

$$\langle S_{AA}(\mathbf{k}) \rangle_{\mathbf{K}} = \frac{1}{N} \sum_{p \in A, q \in A} w_{p,q}^{\mathbf{K}},$$

$$\langle S_{BB}(\mathbf{k}) \rangle_{\mathbf{K}} = \frac{1}{N} \sum_{p \in B, q \in B} w_{p,q}^{\mathbf{K}}, \quad (S4)$$

$$\langle S_{AB}(\mathbf{k}) \rangle_{\mathbf{K}} = \frac{1}{N} \sum_{\substack{(p \in A, q \in B) \text{ or} \\ (p \in B, q \in A)}} w_{p,q}^{\mathbf{K}}.$$

The node weight degree is decomposed in the same way as:

$$w_{p,AA}^{\mathbf{K}} = \begin{cases} \sum_{q \in A} w_{p,q}^{\mathbf{K}} & (p \in A) \\ 0 & (p \in B) \end{cases}$$

$$w_{p,BB}^{\mathbf{K}} = \begin{cases} 0 & (p \in A) \\ \sum_{q \in B} w_{p,q}^{\mathbf{K}} & (p \in B) \end{cases} \quad (S5)$$

$$w_{p,AB}^{\mathbf{K}} = \begin{cases} \sum_{q \in B} w_{p,q}^{\mathbf{K}} & (p \in A) \\ \sum_{q \in A} w_{p,q}^{\mathbf{K}} & (p \in B) \end{cases}$$

The relationship between a structure factor and node degrees also holds for each multipartite network, as follows:

$$\langle S_{AA}(\mathbf{k}) \rangle_{\mathbf{K}} = \frac{1}{N} \sum_{p \in A} w_{p,AA}^{\mathbf{K}}, \quad \langle S_{BB}(\mathbf{k}) \rangle_{\mathbf{K}} = \frac{1}{N} \sum_{p \in B} w_{p,BB}^{\mathbf{K}}, \quad \langle S_{AB}(\mathbf{k}) \rangle_{\mathbf{K}} = \frac{1}{N} \sum_{\substack{(p \in A, q \in B) \text{ or} \\ (p \in B, q \in A)}} w_{p,q}^{\mathbf{K}}. \quad (S6)$$

**Note S3. Form factors of particles**

To extract the effect of particle form factors in edge weights, we start with the Fourier transform of a two-dimensional (2D) isotropic function $V(\mathbf{r})$ obtained as[2]

$$v(\mathbf{k}) = \int V(\mathbf{r})e^{-i\mathbf{k}\cdot\mathbf{r}}d\mathbf{r} = 2\pi\int_0^\infty rV(r)J_0(kr)dr, \qquad (S7)$$

where $r$ and $k$ are norms of $\mathbf{r}$ and $\mathbf{k}$, respectively, and $J_0(x)$ is the zeroth-order Bessel function. We assume that a particle is a constant potential perturbation $V$ bounded inside a circular region of radius $R$. Then, $V(r)$ is represented as

$$V(r) = V\Theta(R-r) = \begin{cases} V & (r \leq R) \\ 0 & (r > R) \end{cases} \qquad (S8)$$

where $\Theta(x)$ is the unit step function. The Fourier transform, $v(\mathbf{k})$, of a particle becomes

$$\begin{aligned}v(\mathbf{k}) &= \frac{2\pi V}{k^2}\int_0^{kR}\tau J_0(\tau)d\tau \\ &= 2\pi VR^2\frac{J_1(kR)}{kR}\end{aligned} \qquad (S9)$$

According to Eq. (S8), we define the form factor of a particle as $f \equiv V\pi R^2$. When applying the point-particle approximation by $R \to 0$, we obtain $v(\mathbf{k}) = f$ from Eq. (S9). Therefore, the Fourier transform $v_p(\mathbf{k})$ of the $p$th particle is expressed as $v_p(\mathbf{k}) = f_p\exp(-i\mathbf{k}\cdot\mathbf{r}_p)$, where $\mathbf{r}_p$ and $f_p$ are the position and the form factor of the $p$th particle, respectively. The average scattering intensity $I_{\text{avg}}(\mathbf{k})$ becomes constant as well, given by $I_{\text{avg}}(\mathbf{k}) = (1/N)\sum_p f_p^2$. Finally, the edge weight is expressed as:

$$w_{p,q}^{\mathbf{K}} = \frac{f_p f_q}{I_{\text{avg}}}\frac{1}{\Omega_{\mathbf{K}}}\int_{\mathbf{K}}\cos(\mathbf{k}\cdot(\mathbf{r}_p-\mathbf{r}_q))d\mathbf{k}. \qquad (S10)$$

One can find that the edge weight has the scattering-strength multiplicative term $f_p f_q/I_{\text{avg}}$. If there are two-phase particles 'A' and 'B', the term $f_p f_q/I_{\text{avg}}$ depends only on $f_B/f_A$, where $f_A$ and $f_B$ are form factors of phase 'A' and 'B' particles, respectively. Therefore, this form factor ratio fully characterizes the inhomogeneity arising from phase-dependent strengths of potential perturbations in our particle modelling. In the main text, we select $\mathbf{K} = \{\mathbf{k} \mid |\mathbf{k}| < k_{\text{th}}\}$, which allows edge weights calculated as

$$\begin{aligned}
w_{p,q}^{\mathbf{K}} &= \frac{f_p f_q}{I_{\text{avg}}} \frac{1}{\pi k_{\text{th}}^2} \int_{k=0}^{k=k_{\text{th}}} \int_0^{2\pi} \cos(k|\mathbf{r}_p - \mathbf{r}_q|\cos\phi) k \, d\phi \, dk \\
&= \frac{f_p f_q}{I_{\text{avg}}} \frac{1}{\pi k_{\text{th}}^2} \times 2\pi \int_0^{k_{\text{th}}} k J_0(k|\mathbf{r}_p - \mathbf{r}_q|) dk \\
&= \frac{f_p f_q}{I_{\text{avg}}} \times \frac{2 J_1(k_{\text{th}}|\mathbf{r}_p - \mathbf{r}_q|)}{k_{\text{th}}|\mathbf{r}_p - \mathbf{r}_q|}.
\end{aligned} \tag{S11}$$

**Note S4. Bipartite networks as material microstructure identifiers**

As shown in Eq. (S10), the term $f_p f_q / I_{avg}$ serves as a scaling factor. Because $f_A^2/I_{avg}$ and $f_B^2/I_{avg}$ are different in general if $f_A \neq f_B$, unipartite network distributions can differ even if two heterogeneous material phases possess an identical arrangement. In contrast, edges of a bipartite network have a common multiplicative factor of $f_A f_B / I_{avg}$, enabling the identification of material microstructures regardless of the heterogeneity in material phases.

To examine the utilization of bipartite networks as material microstructure identifiers, we compare various heterogeneous materials composed of two sets of multiparticles at different phases and with distinct microstructures: two Poisson materials (A and B in Fig. S1a), a Poisson material (A) and a crystal (B) (Fig. S1b), a Poisson material (A) and a more confined Poisson material across radial (B, Fig. S1c) and angular (B, Fig. S1d) axes. When considering the directional bipartite network from B to A ($w_{p \in B, b}^K$), all the degree distributions in Fig. S1 are almost same due to the statistically identical distribution of phase A particles. By contrast, the alteration of phase B microstructures differentiates the degree distributions from the directional bipartite networks from A to B ($w_{p \in A, b}^K$). Notably, while the crystal phase B is reflected in the Bragg-like peak in Fig. S1b, the entire filling ratio and the type of inhomogeneity are reflected in the probability density broadening (Fig. S1a versus Fig. S1c,d) and the enhanced probability density at large $w_{p \in A, b}^K$ (Fig. S1c versus Fig. S1d), respectively.

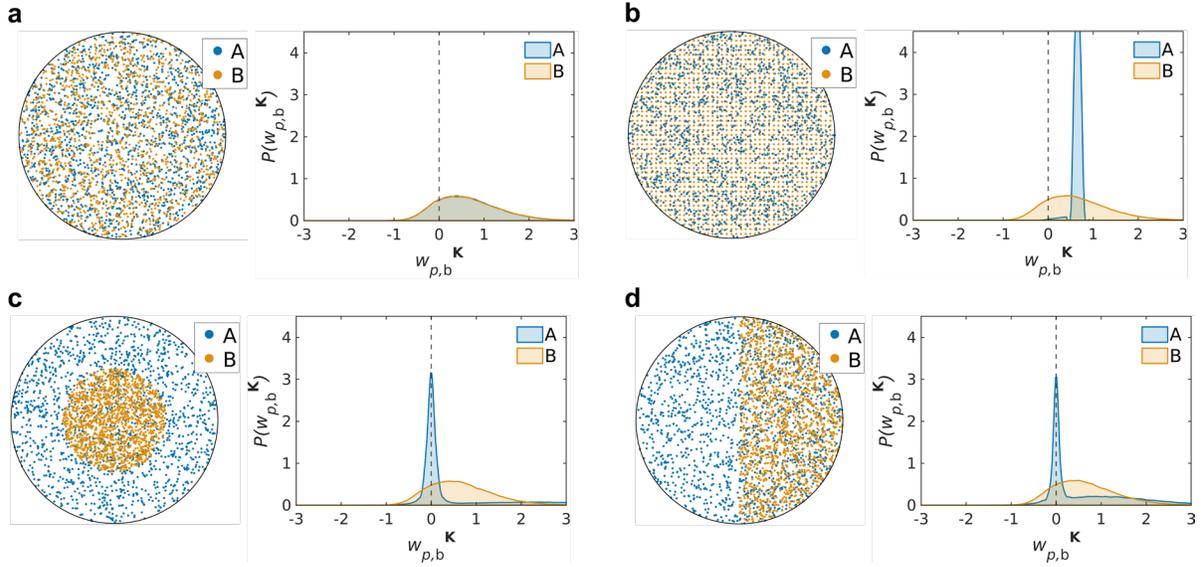

**Fig. S1. Identification of material microstructures using bipartite networks. a-d,** Sample materials and bipartite net degree distributions with distinct microstructures. Two Poisson materials (A and B) (**a**), a Poisson material (A) and a crystal (B) (**b**), a Poisson material (A) and a more confined Poisson material along radial (B) (**c**) and angular (B) (**d**) axes. All distributions are obtained by evaluating 100 realizations, each consisting of 3,000 particles. Phase inhomogeneity is $f_B/f_A = 0.5$, same as Fig. 3-4 of the main text.

**Note S5. Microstructural statistics of finite-size materials**

For comparison, Fig. S2 shows the microstructural statistics of finite materials, including a crystal (Fig. S2a-c), SHU (Fig. S2d-f), and Poisson materials (Fig. S2g-i). While a crystal and SHU exhibit rapid decay of $\sigma^2(R)/\mu(R)$ (Fig. S2b,e) and reveal suppression region of structure factors in the target reciprocal space **K** (Fig. S2c,f), Poisson materials show $\sigma^2(R)/\mu(R) \approx 1$ over length scales (Fig. S2h) and the absence of suppressed region in the structure factor (Fig. S2i). We note that the spatial finiteness of the material leads to nonideal decay ($\sigma^2(R)/\mu(R) \neq 1/R$) in the SHU material.

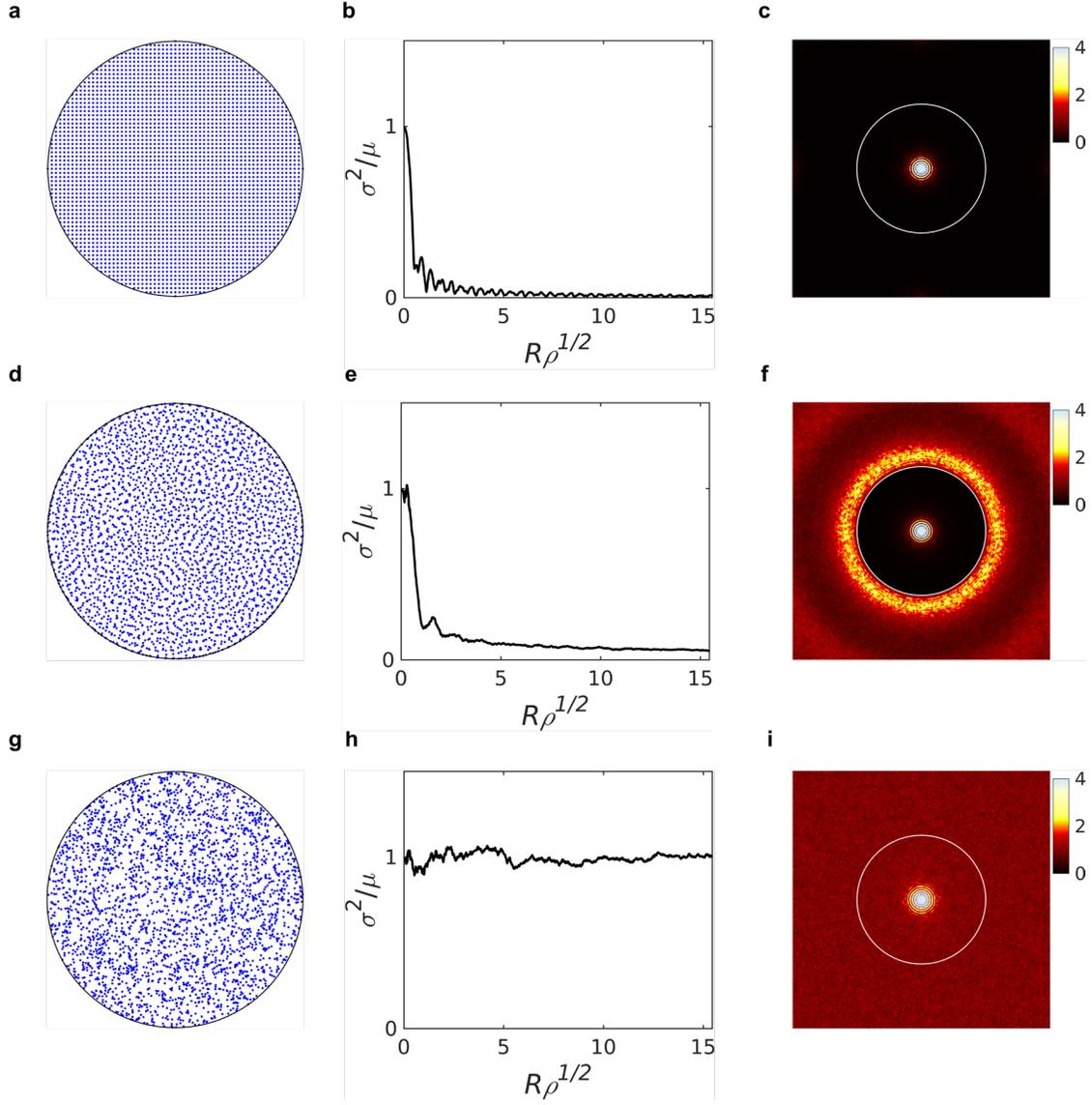

**Fig. S2. Properties of finite-size materials.** Crystal (**a-c**), SHU (**d-f**), and Poisson materials (**g-i**). Sample materials (**a,d,g**), scaled number variances (**b,e,h**), and ensemble-averaged structure factor profile in reciprocal space (**c,f,i**). White circles in (**c,f,i**) denote the target reciprocal space **K** for scattering suppression. Each crystal and SHU materials consist of 3,001 and 3,000 identical particles, respectively. Poisson materials are generated from a Poisson distribution of rate parameter $\lambda = 3,000$, with identical particles. Scaled number variances and ensemble-averaged structure factors are obtained from 100 realizations each.